\documentclass[12pt]{article} % use larger type; default would be 10pt

\usepackage[utf8]{inputenc} % set input encoding (not needed with XeLaTeX)

\usepackage{authblk}

\usepackage{amsmath}
\usepackage{graphicx,psfrag,epsf}
\usepackage{enumerate}
\usepackage{natbib}
\usepackage{url} 
\usepackage{enumitem}

\RequirePackage{amsthm,amsmath,amsfonts}
\usepackage{multirow}
\usepackage[ruled]{algorithm2e}
\usepackage{comment}
\RequirePackage[citecolor=blue,urlcolor=blue]{hyperref}

\usepackage{pgfplots}
\usepackage{pgfplotstable}
\usepackage{here}
\usepackage{float}
\usepackage[font=small]{caption}
\usepackage{subcaption}

\setcitestyle{numbers,square}

\numberwithin{equation}{section}
\newcommand{\R}{\mathbb{R}}
\newcommand{\N}{\mathbb{N}}
\newcommand{\M}{\mathbb{M}}
\newcommand{\X}{\mathbf{X}}
\newcommand{\Y}{\mathbf{Y}}

\newcommand{\x}{\mathbf{x}}

\def\bSig\mathbf{\Sigma}

\renewcommand\P{\mathbb{P}}

\usepackage{xcolor}
\colorlet{lightblue}{blue!10}
\usepackage{amssymb}
\newcounter{exno}
\newcommand{\ex}[1]{\refstepcounter{exno}\label{#1}}

\graphicspath{{.}{./Figures/}}

\begin{document}

\def\spacingset#1{\renewcommand{\baselinestretch}%
{#1}\small\normalsize} \spacingset{1}

%%%%%%%%%%%%%%%%%%%%%%%%%%%%%%%%%%%%%%%%%%%%%%%%%%%%%%%%%%%%%%%%%%%%%%%%%%%%%%

\title{\bf A generative model to synthetize  spatio-temporal dynamics of biomolecules in cells}

\author[1,2,3]{Lisa Balsollier}
\author[1]{Fr\'ed\'eric Lavancier}
\author[2,3]{Jean Salamero}
\author[2,3]{Charles Kervrann}

\affil[1]{{\footnotesize  LMJL, UMR 6629 CNRS, Nantes Universit\'e, F-44322 Nantes, France}}
\affil[2]{{\footnotesize SERPICO Project-Team, Centre Inria de l’Universit\'e de Rennes, F-35042 Rennes Cedex, France}}
\affil[3]{{\footnotesize  Institut Curie, UMR CNRS 144, PSL Research University, Sorbonne Universit\'es, F-75005 Paris, France}}

\date{}
\maketitle

\vspace{-8.5mm}

\begin{abstract}

 Generators of space-time dynamics in bioimaging have become essential to build ground truth datasets for image processing algorithm
evaluation such as biomolecule detectors and trackers, as well as to generate training
datasets for deep learning algorithms. In this contribution, we leverage a stochastic model, called birth-death-move (BDM) point process, in order to generate joint dynamics of biomolecules in cells. This approach is very flexible and allows us to model a system of particles in motion, possibly in interaction, that can each possibly switch from a motion regime (e.g. Brownian) to another (e.g. a directed motion), along with the appearance over time of new trajectories and their death after some lifetime, all of these features possibly depending on the current spatial configuration of all existing particles. We explain how to specify all characteristics of a BDM model, with many practical examples that are relevant for bioimaging applications. Based on real fluorescence microscopy datasets, we finally calibrate our model to mimic the joint dynamics of Langerin and Rab11 proteins near the plasma membrane. We show that the resulting synthetic sequences exhibit comparable features as those observed in real microscopy image  sequences.
\bigskip

\noindent%
{\it Keywords:}  Simulation and Image Synthesis, Intracellular dynamics and Molecular Motion, Fluorescence Microscopy, Birth-death-move process, Spatial Statistics.

\end{abstract}

\newpage

\section{Introduction}
\label{sec:intro}
A long-term goal in fundamental biology is to decipher
the spatio-temporal dynamic coordination and organization of interacting molecules within
molecular complexes at the single cell level.
This includes the characterization of intracellular dynamics, which is essential to a better understanding of fundamental mechanisms like membrane
transport. To that end, dedicated
image analysis methods have been developed to process challenging temporal series of
2D-3D images acquired by fluorescence microscopy, see for instance \cite{kervrann2015}. 

In this context,  mathematical and biophysical models are indispensable  to decode the traffic flows of biomolecules. They constitue crucial prior motion models in most particle tracking procedures \cite{chenouard2014,roudot2017}, and they  are needed to carry out simulations in order to, for instance,  evaluate the performance of image analysis algorithms and train complex models like supervised deep neural networks  \cite{lagardere2020,badoual2021}.
In past years, stochastic models for individual trajectories of particles have been introduced, allowing for Brownian, confined or directed motions with variable velocities within the cell 
 \cite{boulanger2009,bressloff2013,hoze2017,Briane18,pecot2018,Briane19,lagardere2020}, and possibly supported along a cytoskeleton network \cite{lagache2009,kervrann2022}.
However, these modelling approaches hardly represent the collective motion of particles and
global biomolecule trafficking. The latter should ideally  include possible interactions between biomolecules, in particular between different types of proteins, giving raise to the colocalisation phenomena observed  in several applications \cite{Bolte2006,Costes2004,Lagache2015,Lavancier20}. 
It should also be able to generate the lifetime of each trajectory, its spatial distribution and its starting time  during the simulated image sequences.

We propose in this contribution to leverage a tailored  stochastic model introduced  in \cite{Lavancier_LeGuevel}, which is able to generate the aforementioned features. This so-called birth-death-move (BDM) spatial point process is a flexible model for the dynamics of a system of particles (here a group of biomolecules), that move over time, while some new particles may appear in the cell and some existing particles may disappear. 
We further add the possibility for each particle to be marked by a given label, e.g. among different possible labeled proteins and/or different types of motion regimes, and this mark may change over time, e.g. 
 a particle may switch from one regime (e.g. Brownian) to another (e.g. subdiffusive).  This switch of a mark  is sometimes called a ``mutation'' in the literature, but we prefer here to use the term ``transformation'' to avoid misunderstanding with a genuine biological mutation. Moreover the trajectories can be driven by any continuous Markov diffusion model, that includes all models  for individual trajectories previously considered in the literature, and some interactions may be introduced so that colocalization phenomena can be generated.  The intensity of births, that govern the waiting time before the next appearance of a new particle, may depend on the current configuration of particles, and similarly for the intensity of deaths. For instance, we may design that the more biomolecules in the cell there are, the higher the death intensity is, implying a rapid  disappearance. Some spatial effects may also be taken into account, in order to create distinct motion regimes in some regions of the cell, or to encourage some spatial regions for the appearance of a new particle, e.g. nearby some existing particles due to colocalization.

\begin{figure}[h]
     \centering
       \includegraphics[height=4.8cm]{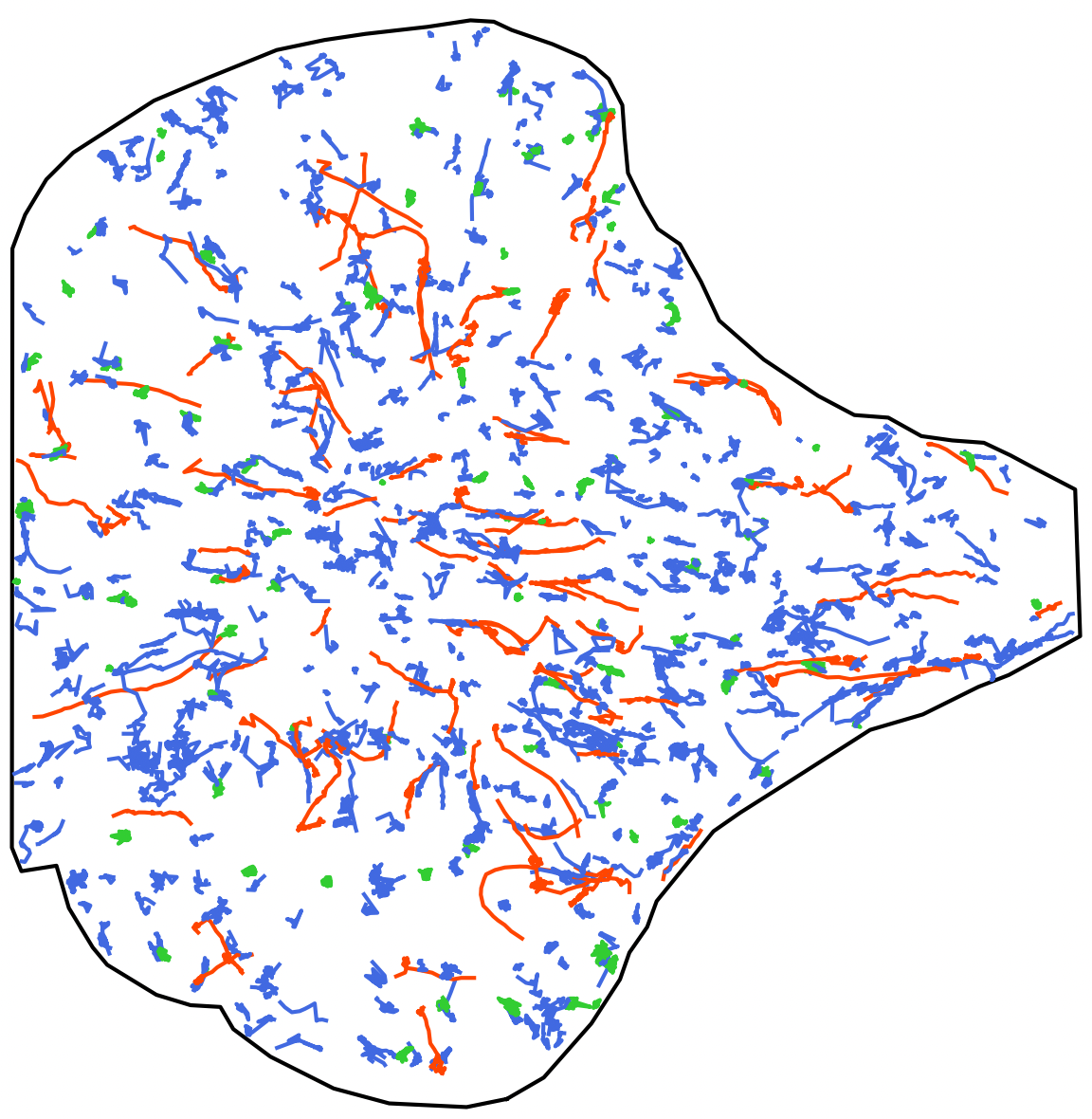}
       \hspace{1cm}
       \includegraphics[height=4.8cm]{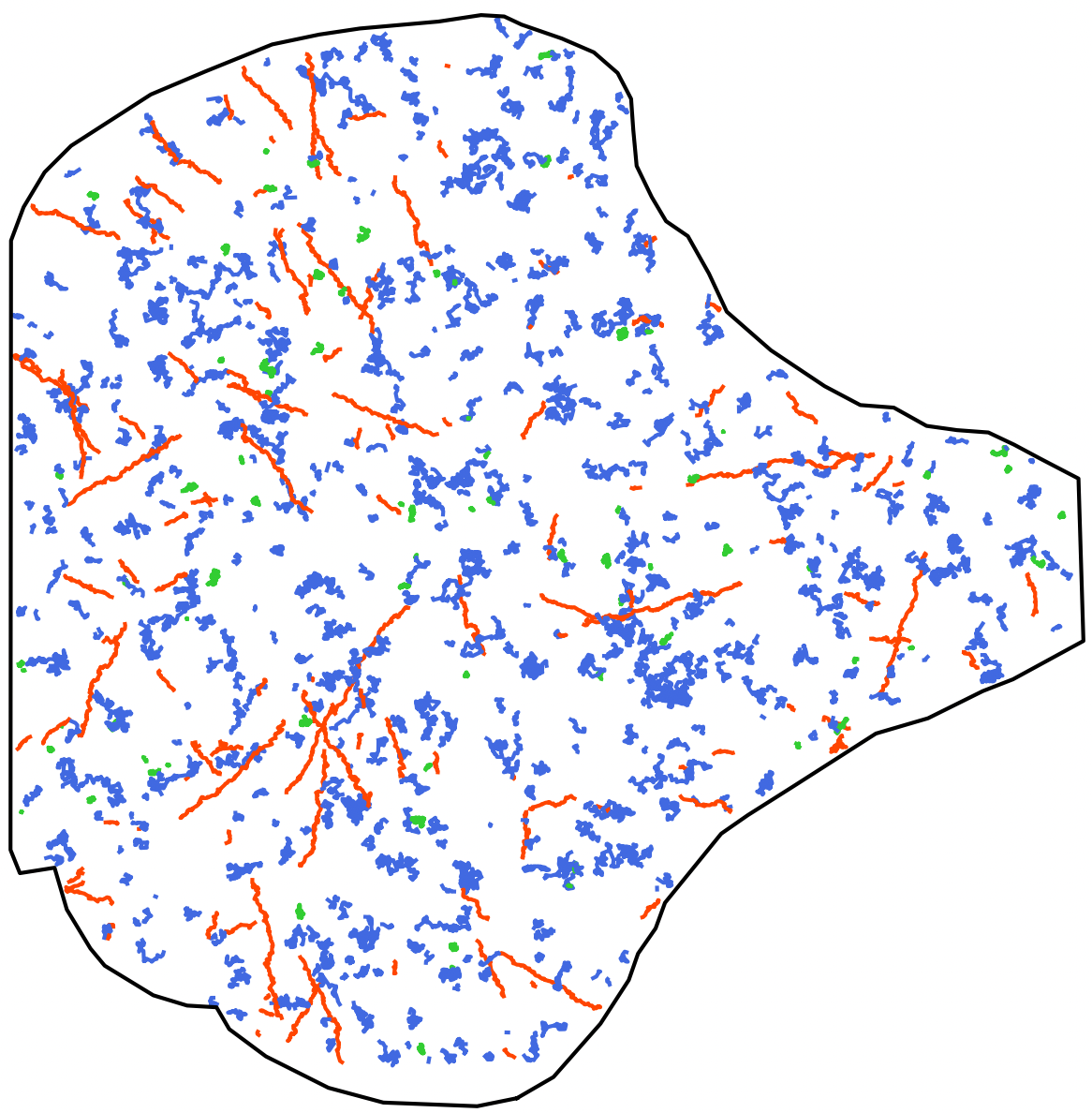}
       \caption{Left: set of all trajectories detected and tracked over a real image sequence of Langerin proteins, colored by their estimated motion regime (Brownian in blue, superdiffusive in red and subdiffusive in green). Right: result from a synthetic sequence generated by our stochastic model.}
       \label{fig-intro}
\end{figure}

The remainder of the paper is organized as follows. In Section~\ref{sec:model}, we give the precise  definition of our stochastic process, and we in particular list all ingredients needed to fully specify the model. An iterative construction  is presented in Section~\ref{sec:algo}, clarifying how the dynamics proceeds, and an effective simulation algorithm is formally detailed in the appendix and made available online. In Section~\ref{sec:examples}, we provide numerous examples for the specifications of the model, that we think are relevant for many real biomolecules dynamics. 
In Section~\ref{sec:appli}, we demonstrate the potential of our generator by focusing on the joint dynamics of Langerin and Rab11 proteins being involved in membrane trafficking. Our approach starts by the inspection of a real dataset in order to choose and calibrate judiciously the different parameters of the BDM model. This dataset consists of a sequence of images acquired by 3D multi-angle TIRF (total internal reflection fluorescence) microscopy technique \cite{Boulanger2014}, depicting the locations of Langerin and Rab11 proteins   close to
the plasma membrane of the cell, specifically over a distance of  1 $\mu$m in the z-axis. After some post-processing, the sequence shows a set of trajectories for both type of proteins, that follow different motion regimes, are spatially distributed within the cell in a specific way, and occur at different periods during the sequence, which is in perfect line with the dynamics of a BDM process. 
The observed trajectories for the Langerin channel are depicted on the leftmost plot of Figure~\ref{fig-intro}. 
 We compute a set of descriptors from this dataset in order to calibrate the parameters of our stochastic process, but also to create some benchmark features for the assessment of our synthetic sequences. Finally, in Section~\ref{sec:simus}, we generate several simulated sequences and show that they exhibit comparable features as those observed in the real sequence. An example of generated trajectories is displayed in the rightmost plot of Figure~\ref{fig-intro}.

Supplementary materials, including the Python code for simulation, the raw data and some further simulated sequences, are available in our online GitHub repository at \url{https://github.com/balsollier-lisa/BDM-generator-for-bioimaging}.

\section{The mathematical model}\label{sec:model}

\subsection{Heuristic and notations}
In order to mimic the dynamics of biomolecules, we consider a multitype birth-death-move process with mutations, denoted by $(\X_{t})_{t\geq0}$.This process is a generalisation of birth-death-move processes, as introduced in \cite{Lavancier_LeGuevel}. In the following, to avoid misunderstanding we rather use the term ``transformation'' instead of ``mutation'', as explained in Section~\ref{sec:intro}. This section describes the spatio-temporal dynamics of $(\X_{t})_{t\geq0}$ and introduces some notation.

At each time $t$, $\X_t$ is a collection of particles located in a bounded set $\Lambda$ of $\R^d$. 
Each particle is assigned a mark that represents a certain feature. We denote by $\M$ the collection of possible marks.
Through time, the particles move  (possibly depending on their associated mark and  in interaction with each other) and three sudden changes may occur, that we call ``jumps'':
\begin{enumerate}
\item a ``birth'': a new particle, assigned with a mark, may appear;
\item  a ``death'': an existing particle may disappear;
\item a ``transformation'': the mark of an existing particle may change.
\end{enumerate}

\medskip

\noindent {\bf Example}: In our biological application treated in Section~\ref{sec:appli}, $\Lambda$ represents a cell in dimension $d=2$ or $d=3$. We observe inside this cell two types of particles, associated to Langerin and Rab11 proteins, and  each of them moves according to three different possible regimes: Brownian, superdiffusive or subdiffusive. For this example, each particle is therefore marked out of 6 possibilities, whether it is associated to Langerin (L) or Rab11 (R), and depending on its motion regime (1 to 3), so that $\M=\left\{ (L,1),(L,2),(L,3), (R,1),(R,2),(R,3)\right\}$. Through time, each particle moves independently of the others according to its motion regime and eventually a new particle may appear, an existing one may disappear, and the motion regime of some particles may change. 

\medskip

We denote by $n(\X_t)$ the number of particles at time $t$. Each particle $x_i\in \X_t$, for $i=1,\dots,n(\X_t)$, is decomposed as $x_i=(z_i,m_i)$ where $z_i\in\Lambda$ stands for its position while $m_i\in\M$ denotes its mark. Accordingly we have $\X_t\in  (\Lambda\times \M)^{n(\X_t)}$. (Strictly speaking the ordering of particles in $\X_t$ does not matter,  because  any permutation of particles  leads to the same collection of particles. We choose in this paper to bypass this nuance and use the same notation as if $\X_t$ was a vector a particles, even it is actually a set of particles.)   Since the number of particles changes over time, the stochastic process $(\X_{t})_{t\geq0}$ takes its values in the space 
$$E=\bigcup_{n\in\N}  (\Lambda\times \M)^n.$$
To stress the fact that $\X_t$ is not a simple value but encodes the positions and marks of a system of particles, we will say that this system at time $t$ is in {\it configuration} $\X_t$. 

\medskip

To fully specify the dynamics of $(\X_t)_{t\geq 0}$, we need the following ingredients: 
 \begin{enumerate}
 \item A system of equations $(Move)$ that rules the way each particle of $\X_t$ moves continuously between two jumps. We will typically consider a system of stochastic differential equations acting on the position of each particle, possibly depending on their associated mark and  in interaction with the other particles;

 \item Three continuous bounded functions $ \beta$, $ \delta$ and $ \tau$ from $E$ to $[0,\infty)$, called  birth, death and transformation intensity functions respectively, that govern the waiting times before a new birth, a new death and a new transformation. At each time $t$, we may interpret $\beta(\X_t)dt$ as the probability that a birth occurs in the interval $[t, t+dt]$, given that the system of particles is in the configuration  $\X_t$,  and similarly for $\delta$ and $\tau$. 
 
 \item Three transition probability functions that indicate how each jump occurs: %Three transition probability functions that indicate how each jump occurs. Assume that the particles are in the  configuration $\x\in E$ at the jump time, whether this jump is a birth, a death or a transformation. Then we introduce:
 \begin{itemize}
 \item $p^\beta((z,m)|\x)$: probability density function that the birth occurs at the position $z$ with the mark $m$, given that there is a birth and that the system of particles is in configuration $\x$ at the birth time;
 \item $p^\delta(x_i|\x)$, for $x_i\in \x$: probability that the death concerns the particle $x_i$ in $\x$, given that there is a death and that the system of particles is in configuration $\x$ at the death time;
 \item $p^\tau((z_i,m_i),m|\x)$, for $(z_i,m_i)\in \x$: probability that the particle $x_i=(z_i,m_i)$ in $\x$ changes its mark and that this transformation leads to the new mark $m\neq m_i$, given that there is a transformation  and that the system of particles is in configuration $\x$ at the transformation time.

 \end{itemize}

\end{enumerate}

We provide in Section~\ref{sec:examples} some examples for the choice of these characteristics. 
Finally, we will denote by $T_1,T_2,\dots$ the jump times of the process and we agree that $T_0=0$.

\subsection{Algorithmic construction}\label{sec:algo}

Assume we are given the characteristics of the process $(\X_t)_{t\geq 0}$ as introduced in the previous section, that are the system of equations $(Move)$, the intensity functions $\beta$, $\delta$, $\tau$, and the transition probability functions $p^\beta$, $p^\delta$, $p^\tau$. 
Then, starting from an initial configuration $\X_{0}$ at time $T_0=0$, we  construct iteratively the process in the time interval  $[0,T]$ as follows.  Here we set $\alpha=\beta+\delta+\tau$ to be the total intensity of jumps. 
\begin{enumerate}
    \item Generate $n(\X_0)$ continuous trajectories as solutions of $(Move)$ in the interval $[0,T-T_0]$, given the initial conditions $\X_0$. Denote $(\Y_t)_{t\in[0,T-T_0]}$ these trajectories. 
    \item By flipping a coin, test whether the jump time $T_1$ occurs after $T$ (this is with probability $p$) or before $T$ (this is with probability $1-p$), where
    $$p=\exp\left(-\displaystyle\int_{0}^{T-T_0}\alpha(\Y_u)d u\right).$$
    \begin{itemize}
      \item If $T_1>T$, then $\X_t= \Y_{t-T_0}$ for all $t\in [T_0,T]$, which completes the simulation. 
      \item Otherwise, we continue by generating $T_1$ in $[T_0,T]$ and the associated jump as in the following.
      \end{itemize}
    \item Generate $T_1-T_0$, given that $T_1<T$, according to the probability distribution
    $$\P(T_1-T_0<t| T_1<T)=\frac 1 {1-p}\left(1-\exp\left(-\displaystyle\int_0^t \alpha(\Y_u)d u\right)\right),\quad 0<t<T-T_0.$$
The process until the time $T_{1}$ is then given by the generated trajectories, i.e.
     $$(\X_t)_{T_0\leq t< T_1}=(\Y_{t-T_0})_{T_0\leq  t< T_1}.$$
     
   \item Draw which kind of jump occurs at $T_1$ (we denote  by $\X_{T_1^-}$ the configuration of the process just before the jump, which is $\Y_{T_1-T_0}$ by continuity of $(\Y_t)$):
   \begin{itemize}
   \item this is a birth with probability $\beta(\X_{T_{1}^-})/ \alpha(\X_{T_{1}^-})$; 
   \item this is a death with probability $\delta(\X_{T_{1}^-})/ \alpha(\X_{T_{1}^-})$;
   \item this is a transformation with probability $\tau(\X_{T_{1}^-})/ \alpha(\X_{T_{1}^-})$.
   
   \end{itemize}
     
   %  \item We generate $C$ according to the law with three possibilities $\left\{ B,D,M\right\}$ of probability $$\left\{  \dfrac{ \beta(X_{T_{1}^-})}{ (\beta+ \delta+ \tau)(X_{T_{1}^-})}, \dfrac{ \delta(X_{T_{1}^-})}{ (\beta+ \delta+ \tau)(X_{T_{1}^-})}, \dfrac{ \tau(X_{T_{1}^-})}{ (\beta+ \delta+ \tau)(X_{T_{1}^-})}\right\}$$
    \item Generate the jump at $t=T_1$ to get $\X_{T_{1}}$ as follows:
     \begin{itemize}
   \item if this is a birth, generate the new particle $x=(z,m)$ according to the probability density function $p^\beta((z,m)|\X_{T_{1}^-})$. Then set $\X_{T_1}=\X_{T_{1}^-} \cup (z,m)$;
 \item  if this is a death, draw which particle $x_i\in \X_{T_{1}^-}$ to delete according to the probability $p^\delta(x_i|\X_{T_{1}^-})$, for $i=1,\dots,n(\X_{T_{1}^-})$. Then  set $\X_{T_1}=\X_{T_{1}^-} \setminus x_i$;
 \item if this is a transformation, draw which particle $x_i=(z_i,m_i)\in \X_{T_{1}^-}$ is transformed and generate its transformation  according to $p^\tau((z_i,m_i),m|\X_{T_{1}^-})$, for $i=1,\dots,n(\X_{T_{1}^-})$. Then  set $\X_{T_1}=(\X_{T_{1}^-} \setminus (z_i,m_i)) \cup (z_i,m)$.
 \end{itemize}

   \item Back to step 1 with $T_0\leftarrow T_1$ and $\X_0\leftarrow  \X_{T_1}$ in order to generate the new trajectories starting from $\X_{T_1}$ and the next jump time $T_2$, and so on. 
%       \item Given $X_{T_1}$, we generate the solution $\Y^1$ of $(M_{X_{T_1}})$ 
   % \item \dots and so on \dots
\end{enumerate}

In the first step of the above construction,  the trajectories are generated up to the final time $T$. It is however very likely that the next jump occurs much before $T$ so that it would be sufficient and computationally more efficient to generate these trajectories on a shorter time interval. We provide in Appendix a formal algorithm of simulation of $\X_t$ for $t\in[0,T]$, following the above construction and including the latter idea. This algorithm has been implemented in Python and is available in our GitHub repository. 

From a theoretical side, note that the specific exponential form of the probability distribution of the inter-jump waiting time in step 3 is necessary to imply the interpretation of $\beta$, $\delta$ and $\tau$ explained in the previous section. %: the probability that a birth occurs in the time interval $[t,t+dt]$ given that the particles are in configuration $\X_t$ is $\beta(\X_t)dt$, and similarly for $\delta$ and $\tau$. 
This exponential form also implies that $(\X_t)_{t\geq 0}$ is a Markov process, meaning that its future dynamics only depends on its present configuration. We refer to \cite{Lavancier_LeGuevel} for more details about these theoretical aspects.

\subsection{Exemplified specifications of the model}\label{sec:examples}

\subsubsection{The inter-jumps motion}

Recall that during an inter-jump period, the process $(\X_t)$ has a constant cardinality $n=n(\X_t)$ and the marks of all its particles remain constant. We denote by $(\Y_t)$ a system of $n$ such particles $(z_{i,t},m_i)$, for $i=1,\dots,n$, where $z_{i,t}\in\Lambda$ represents the position of the $i$-th particle at time $t$ and $m_i\in\M$ is its constant mark, that is
$$\Y_t=((z_{1,t},m_1),\cdots,(z_{n,t},m_n)).$$
In agreement with the construction of the previous section, the inter-jump trajectory of each particle of $(\X_t)$ will coincide with the $n$ trajectories of $(\Y_t)$ during this period. 

As a general example, we assume that $(\Y_t)$ follows the following system of stochastic differential equation, starting at $t=0$ at the configuration $\mathbf y_0\in (\Lambda\times \M)^n$,
\begin{align*}
    (Move):\quad \left\{
    \begin{array}{ll}
    d z_{i,t}=v_i(t,\Y_{t})d t +\sigma_i(t,\Y_t)d B_{i,t},& t\geq 0, \quad i=1,\dots,n,\\
    \Y_{0}=\mathbf y_0,& 
    \end{array}
   \right.
\end{align*} 
where the drift functions $v_i$ take their values in $\R^d$,  the diffusion $\sigma_i$ are non-negative functions, and $(B_{i,t})$, $i=1,\dots,n$, are $n$ independent standard Brownian motions in $\R^d$. Here,  $\mathbf y_0$, $v_i$ and $\sigma_i$ are free parameters to be chosen. 

Some conditions on the drift and diffusion functions are necessary to ensure the existence and unicity of the solution of $(Move)$. This holds  for instance if these functions are Lipschitz \citep{oksendal}, a condition met for the following examples. In addition, since each particle is supposed to evolve in the bounded set $\Lambda$ of $\R^d$, we need in practice to force the trajectories of $(Move)$ to stay in $\Lambda$. This may be achieved by reflecting the trajectories at the boundary of $\Lambda$. 

In its general form,  $(Move)$ allows the motion of each particle to depend on its mark, but also on the position and mark of the other particles (that are part of  $\Y_t$). We detail several examples below, that may be realistic for biological applications.

\medskip

%{\it Example:} If $\sigma=0$, then (Move) becomes a system of first-order ordinary differential equations.

\ex{brownian} \noindent \textbf{Example \ref{brownian}}\textit{(Brownian motions)}: 
 If $v_i(t,\Y_t)=0$ and $\sigma_i(t,\Y_t)=\sigma$ (for $\sigma>0$) is constant, then each particle follows a Brownian motion with the same diffusion coefficient $\sigma$, independently of the other particles.

\medskip

\ex{spatial diffusion} \noindent \textbf{Example \ref{spatial diffusion}}\textit{(spatially varying diffusion coefficients)}: 
 If $v_i(t,\Y_t)=0$ and $\sigma_i(t,\Y_t)=\sigma_{m_i}(z_{i,t})$, where $\sigma_{m_i}$ is a positive fonction defined on $\Lambda$, then each particle follows an independent diffusive motion, where the diffusive coefficient depends on the associated mark and may vary in space. For instance, assume that $\Lambda=\Lambda_1\cup\Lambda_2$ with $\Lambda_1\cap\Lambda_2=\varnothing$ and that  for $m\in \M$,
$$\sigma_{m}(z)=\begin{cases} \sigma_{1,m} & \text{if}\quad  z\in\Lambda_1, \\ \sigma_{2,m} & \text{if}\quad  z\in\Lambda_2,\end{cases}$$
where  $\sigma_{1,m}>0$, $\sigma_{2,m}>0$. Then each particle with mark $m$ follows locally in $\Lambda_1$ a Brownian motion with diffusion coefficient 
  $\sigma_{1,m}$ and  locally in $\Lambda_2$ a Brownian motion with diffusion coefficient 
  $\sigma_{2,m}$. Note that as such, $\sigma_m$ is not Lipschitz and it needs to be smooth so as to fit the theoretical setting. This may be achieved by taking the convolution of $\sigma_m$ by a bump function. 
  
  \medskip

\ex{super sub} \noindent \textbf{Example \ref{super sub}}\textit{(superdiffusive and subdiffusive motions)}: 
If $v_i(t,\Y_t)=v_{m_i}(z_{i,t})$ and $\sigma_i(t,\Y_t)=\sigma_{m_i}$, where $v_{m_i}$ is defined on $\Lambda$ and $\sigma_{m_i}>0$, then each particle evolves independently of each other with a drift and a diffusion coefficient that depend on its mark. This example includes the superdiffusive case considered in 
\cite{Briane18} when $v_{m_i}(z)=v_{m_i}$ is a constant drift. It also includes the Ornstein-Uhlenbeck dynamics, also considered in \cite{Briane18}, when $v_{m_i}(z_{i,t})= -\lambda_{m_i}(z_{i,t}-z_{i,0})$, where $\lambda_{m_i}>0$ can be interpreted as a force of attraction towards the initial position $z_{i,0}$, leading to a confined trajectory.

\medskip

\ex{interacting} \noindent \textbf{Example \ref{interacting}}\textit{(interacting particles)}: 
 In this example, we show how we can include interactions between the particles through a Langevin dynamics. To do so, we introduce, for $m,m'\in \M$, pairwise interaction functions $\Phi_{m,m'}$, as considered in statistical physics:  For  $r>0$,  $\Phi_{m,m'}(r)$ represents the pairwise interaction between a particle with mark $m$ and a particle with mark $m'$ at a distance $r$ apart. If $\Phi_{m,m'}(r)=0$, there is no interaction, if  $\Phi_{m,m'}(r)>0$ there is inhibition between the two particles at distance $r$, and if  $\Phi_{m,m'}(r)<0$ there is attraction. Examples of inhibitive  interaction functions can be found in \cite{manent}. The (overdamped) Langevin dynamics associated to these interactions reads as $(Move)$ with $\sigma_i(t,\Y_t)=\sigma_{m_i}$, $\sigma_{m_i}>0$, and 
$$v_i(t,\Y_t)=-\sum_{j\neq i} \nabla \Phi_{m_i,m_j}(\|z_{i,t}-z_{j,t}\|),$$
where $\nabla$ denotes the Gradient operator. 
Accordingly, each particle moves in a direction that tend to decrease the value of the pairwise interaction function with the other particles. 
%In the long run, such system of particles is spatially distributed as a Boltzmann (or Gibbs) distribution on $\Lambda$, meaning that the probability density function for the locations of the $n$ particles is proportional to $\exp(-\sum_{j\neq i} \Phi_{m_i,m_j}(\|z_{i,t}-z_{j,t}\|))$.

\medskip

\ex{coloc move} \noindent \textbf{Example \ref{coloc move}}\textit{(colocalized particles)}: 
Assume that some particles, say with mark $m$, are thought to be  colocalized with particles having the mark $\tilde m$. This means that we expect the former to be localised nearby the latter and to follow approximately the same motion. Specifically, to let the particle $i$ with mark $m$ be colocalized with the particle $j$ with mark $\tilde m$, we may simply define $z_{i,t}=z_{j,t}+\sigma_i B_{i,t}$, $t>0$,  where $B_{i,t}$ is a standard Brownian motion in $\R^d$ representing the deviation of the trajectory  $i$ around the trajectory  $j$, and $\sigma_i>0$ quantifies the strength of this deviation. Here $z_{j,t}$ may be defined as in the previous examples, for instance as the typical trajectory of a particle with mark $\tilde m$.

\subsubsection{The intensity functions}\label{sec:intensities}

Recall that the intensity functions $\beta$, $\delta$ and $\tau$ rule the waiting times until the next birth, death and transformation, respectively. Heuristically, the probability that a birth occurs in the time interval $[t,t+dt]$ given that the particles are in configuration $\X_t$ is $\beta(\X_t)dt$, and similarly for $\delta$ and $\tau$. 
As a consequence these probabilities may evolve over time according to the configuration of particles, making for instance a death more likely to happen when there are many particles or a high concentration of them in some  region, due to competition. We provide some natural examples below. For each example, any of $\beta$, $\delta$ or $\tau$ can be set similarly, even if we focus only on one of them.

\medskip

\ex{constant intensity} \noindent \textbf{Example \ref{constant intensity}}\textit{(constant intensities)}: 
The simplest situation is when the intensity functions are constant, for instance $\beta(\X_t)=\beta$ with $\beta>0$. Then births appear at a constant rate and we can expect that in average $\beta\times(s-t)$ new particles appear during the interval $[s,t]$.

\medskip

\ex{cardinality intensity} \noindent \textbf{Example \ref{cardinality intensity}}\textit{(intensities depending on the cardinality)}: 
If $\delta(\X_t)=\delta n(\X_t)$, with $\delta>0$, then the more particles there are, the more deaths we observe. This is a natural situation when each particle is thought to have a constant death rate $\delta$, so that the total death intensity for the system of particles at time $t$ is just the sum of them, that is  $\delta n(\X_t)$. 

\medskip

\ex{spatial intensity} \noindent \textbf{Example \ref{spatial intensity}}\textit{(spatially varying intensities)}: 
Assume that the mark of a particle (say its motion regime) has more chance to change in some region of $\Lambda$ than another, then the transformation intensity $\tau$ may reflect this dependency. Let for instance $\Lambda=\Lambda_1\cup\Lambda_2$ with $\Lambda_1\cap\Lambda_2=\varnothing$ and define $\tau(\X_t)=\tau_1 n_1(\X_t) + \tau_2 n_2(\X_t)$ where $\tau_1<\tau_2$ and $n_1(\X_t)$ (resp. $n_2(\X_t)$) denotes the number of particles in $\Lambda_1$ (resp. in $\Lambda_2$). Then for a given cardinality $n(\X_t)$, the more proportion of particles  in $\Lambda_2$, the more transformations happen. Note that in order to be rigorous, we should consider a continuous version of $\tau$, which can be achieved by convolution with a bump function. 
%\sum_{i=1}^{n(\X_t)} \mathbb 1_{x_i\in 

\medskip

\ex{coloc intensity} \noindent \textbf{Example \ref{coloc intensity}}\textit{(transformation due to colocalization)}: 
Assume that some $m$-particles (that are the particles with mark $m$) can be colocalized with some $\tilde m$-particles. Assume in addition that the particles are assigned a second mark  that encodes their motion regime (e.g. diffuse, subdiffusive or superdiffusive). Eventually, during the dynamics of particles,  a  non-colocolized $m$-particle may become colocolized with a $\tilde m$-particle, meaning that it becomes $D$-close to a $\tilde m$-particle, where $D$ is some prescribed colocalization distance. If so, we may expect that the motion regime of the $m$-particle becomes similar as the $\tilde m$-particle, so that a transformation must occur. Let $n_{D,m,\tilde m}(\X_t)$ be the number of $D$-close pairs of particles with marks $m$ and $\tilde m$, whose motion regimes are different. Then we may define $\tau(\X_t)=\tau \,n_{D,m,\tilde m}(\X_t)$, for some $\tau>0$, so that a transformation (of motion regime here) is very likely to occur when the aforementioned situation happen. Note that  if $\tau$ is large, such transformation will quickly happen as soon as $n_{D,m,\tilde m}(\X_t)=1$, and so $n_{D,m,\tilde m}(\X_t)>1$ will be unlikely to be observed. Here again a smooth version of $\tau$ can be introduced by convolution to ensure its continuity.

\subsubsection{The transition probability functions}

We detail examples for the three possible transitions, in order below: births, deaths and transformations. 

For the births, remember that $p^\beta((z,m)|\x)$ denotes the probability density function (pdf) that a particle appears at the position $z$ with the mark $m$, given that the system of particles are in configuration $\x$. To set this probability, two approaches are possible: 
\begin{enumerate}
\item First drawing the mark $m$ of the new particle with respect to some probability $p^\beta(m|\x)$, then the position of the new particle given its mark according to some pdf $p^\beta(z|\x,m)$. This leads to the decomposition $p^\beta((z,m)|\x)=p^\beta(m|\x)p^\beta(z|\x,m)$.
\item First generating the position of the new particle with respect to some pdf $p^\beta(z|\x)$, then its mark $m$ given the position with probability $p^\beta(m|\x,z)$. This leads to the decomposition $p^\beta((z,m)|\x)=p^\beta(z|\x)p^\beta(m|\x,z)$.
\end{enumerate}

\medskip

\ex{unif births} \noindent \textbf{Example \ref{unif births}}\textit{(uniform births)}: 
 This is the simple example where the births do not depend on the environment, are uniform in space and the marks are drawn with respect to some prescribed probabilities $p_m$, where $\sum_{m\in\M} p_m=1$. The two above approaches then coincide with $p^\beta(m|\x)=p^\beta(m|\x,z)=p_m$
 and $p^\beta(z|\x,m)= p^\beta(z|\x) = 1/|\Lambda|$ for $z\in\Lambda$.

\medskip

\ex{coloc births} \noindent \textbf{Example \ref{coloc births}}\textit{(colocalized births)}: 
 We adopt here the first approach above. We first draw the marks independently of the environment by setting $p^\beta(m|\x)=p_m$ with $\sum_{m\in\M} p_m=1$, as in the previous example. Second, in order to generate the position of a new $m$-particle, thought to be colocalized with the $\tilde m$-particles, we may use a mixture of isotropic normal distribution, centered at each $\tilde m$-particle, with deviation $\sigma>0$. Denoting by $\tilde n(\x)$ the number of $\tilde m$-particles in $\x$ and $\tilde z_i$ their positions ($i=1,\dots,\tilde n(\x)$), this means that 
\begin{equation}\label{gaussian mix}
p^\beta(z|\x,m)= \frac 1 {\tilde n(\x)} \sum_{i=1}^{\tilde n(\x)} \frac 1 {(\sigma\sqrt{2\pi})^d} \exp\left(-\frac{\|z-\tilde z_i\|}{2\sigma^2}\right).\end{equation}
Note that to be rigorous $p^\beta(z|\x,m)$ should be restricted to $\Lambda$ with a proper normalisation, otherwise some particles might be generated outside $\Lambda$. We omit these details. 
%Note that with this specification, some positions might be generated outside $\Lambda$ and to be rigorous $p^\beta(z|\x,m)$ should be restricted to $\Lambda$ with a proper normalisation. We omit these details. 

\medskip

\ex{spatial births} \noindent \textbf{Example \ref{spatial births}}\textit{(spatially dependent new marks)}: 
 We may adopt the second approach by first generating a uniform position for the new particle, i.e. $p^\beta(z|\x)= 1/|\Lambda|$ for $z\in\Lambda$, and second by drawing the mark according to the generated position. Let for instance $\Lambda=\Lambda_1\cup\Lambda_2$ with $\Lambda_1\cap\Lambda_2=\varnothing$ and set 
$$p^\beta(m|\x,z)=\begin{cases} p_{1,m} & \text{if}\quad  z\in\Lambda_1, \\ p_{2,m} & \text{if}\quad  z\in\Lambda_2,\end{cases}$$
where $\sum_{m\in\M} p_{1,m}=\sum_{m\in\M} p_{2,m}=1$. Then depending on the position, the distribution of the marks may be different. 

\bigskip

We  now focus on the death transition, namely  the probability $p^\delta(x_i|\x)$, for $x_i\in \x$,  that the particle $x_i$ in $\x$ disappears when there is a death. 

\medskip

\ex{unif deaths} \noindent \textbf{Example \ref{unif deaths}}\textit{(uniform deaths)}: 
The simplest example is when a death occurs uniformly over the existing particles, that is $p^\delta(x_i|\x)=1/n(\x)$ for $i=1,\dots,n(\x)$.

\medskip

\ex{competition deaths} \noindent \textbf{Example \ref{competition deaths}}\textit{(deaths due to competition)}: 
We may imagine that, due to competition,  a particle is more likely to disappear if there are too many neighbours around it. Let $n_D(x_i)$ be the number of neighbouring particles around $x_i$ within distance $D>0$. To take into account the competition at distance $D>0$, we may define $
p^\delta(x_i|\x)=n_D(x_i)/\sum_{j=1}^{n(\x)} n_D(x_j)$. Similarly, if relevant, we may count the number of neighbours of a certain mark only.

\bigskip

Finally, we focus on $p^\tau((z_i,m_i),m|\x)$, for $(z_i,m_i)\in \x$, which is the probability that the particle $x_i=(z_i,m_i)$ in $\x$ changes its mark from $m_i$ to $m\neq m_i$,  when a transformation happens. Similarly as for the birth transition probability, it is natural to decompose this probability as 
$$p^\tau((z_i,m_i),m|\x)=p^\tau((z_i,m_i)|\x)p^\tau(m|\x,(z_i,m_i)),$$ 
where $p^\tau((z_i,m_i)|\x)$ represents the probability to choose the particle $(z_i,m_i)$ in the configuration  $\x$, in order to change its mark, and $p^\tau(m|\x,(z_i,m_i))$ is the probability to choose the new mark $m$ given that the transformed particle is located at $z_i$ with mark $m_i$.

\medskip

\ex{indep transfo} \noindent \textbf{Example \ref{indep transfo}}\textit{(transformations independent on the environment)}: 
A typical situation is when the particle to transform is drawn uniformly over the existing particles, that is $p^\tau((z_i,m_i)|\x)=1/n(\x)$, and the transformation is carried out independently on the environment, according to a transition matrix with entries $p_{m,m'}\geq 0$, $m,m'\in\M$, representing the probability to be transformed from mark $m$ to mark $m'$. Here, for all $m\in \M$, we assume $p_{m,m}=0$ in order to ensure a genuine transformation, and of course $\sum_{m'\in\M}p_{m,m'}=1$. With this formalism, we thus have $p^\tau(m|\x,(z_i,m_i))=p_{m_i,m}$. 

\medskip

\ex{spatial transfo} \noindent \textbf{Example \ref{spatial transfo}}\textit{(spatially dependent transformations)}: 
To make the previous example spatially dependent,
introduce $p_m(z)$, a pdf in $\Lambda$ representing the locations in $\Lambda$ where a particle with mark $m$ is more or less likely to be transformed. Then we may set 
$$p^\tau((z_i,m_i)|\x)=\frac{n_i(\x)}{n(\x)}  \frac{p_{m_i}(z_i)}{\sum_{j=1}^{n_i(\x)} p_{m_i}(z_j^{(m_i)})} ,$$ 
where $n_i(\x)$ denotes the number of particles with marks $m_i$ in $\x$ and $z_j^{(m_i)}$, $j=1,\dots,n_i(\x)$, their positions. In this expression $n_i(\x)/n(\x)$ is a weight accounting for the prevalence of mark $m_i$ in $\x$ and the sum in the denominator is a normalisation so that the probabilities sum to 1. Note that if $p_m(z)$ is the uniform pdf on $\Lambda$, then we recover the uniform distribution $p^\tau((z_i,m_i)|\x)=1/n(\x)$. Furthermore, once the particle is chosen as above, we may apply a spatially dependent transformation as follows. Let  $\Lambda=\Lambda_1\cup\Lambda_2$ with $\Lambda_1\cap\Lambda_2=\varnothing$ and let two different transition matrices with respective entries $p^{(1)}_{m,m'}$ and $p^{(2)}_{m,m'}$, for $m,m'\in\M$. Then we may set
 $$p^\tau(m|\x,(z_i,m_i))=\begin{cases} p^{(1)}_{m_i,m} & \text{if}\quad  z_i\in\Lambda_1, \\  p^{(2)}_{m_i,m} & \text{if}\quad  z_i\in\Lambda_2.\end{cases}$$
Accordingly, the transformation does not follow the same distribution,  whether the chosen particle to be transformed is located in $\Lambda_1$ or $\Lambda_2$.

\medskip

\ex{coloc transfo} \noindent \textbf{Example \ref{coloc transfo}}\textit{(transformation due to colocalization)}: 
Assume that we are in the same situation as in Example~\ref{coloc intensity} where $m$-particles can be colocalized to $\tilde m$-particles. We assume like in this example that a transformation occurs if $n_{D,m,\tilde m}(\X_t)\geq 1$, where $n_{D,m,\tilde m}(\X_t)$ denotes  the number of $D$-close pairs of particles with marks $m$ and $\tilde m$, whose motion regimes are different. Then, when a transformation happens, we may choose the $m$-particle to be transformed uniformly over those $m$-particles that are $D$-close to a $\tilde m$-particle with a different motion regime. Then the transformation makes the motion regime of the selected $m$-particle similar as the motion regime of its closest $\tilde m$-particle.

\section{Application to the joint dynamics of Langerin/Rab11 proteins}\label{sec:appli}
\subsection{Description of the dataset}\label{sec:data}

\begin{figure}
  \centering
\begin{tabular}{ccc}   
     \includegraphics[height=7cm]{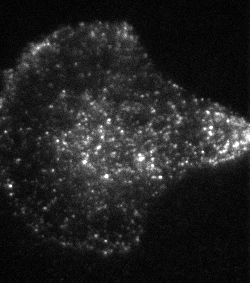} $
  \hspace{30pt} $
         \includegraphics[height=7cm]{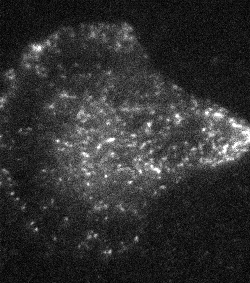}  \\
     \includegraphics[height=7cm]{traj.png}$
  \hspace{30pt} $
     \includegraphics[height=7cm]{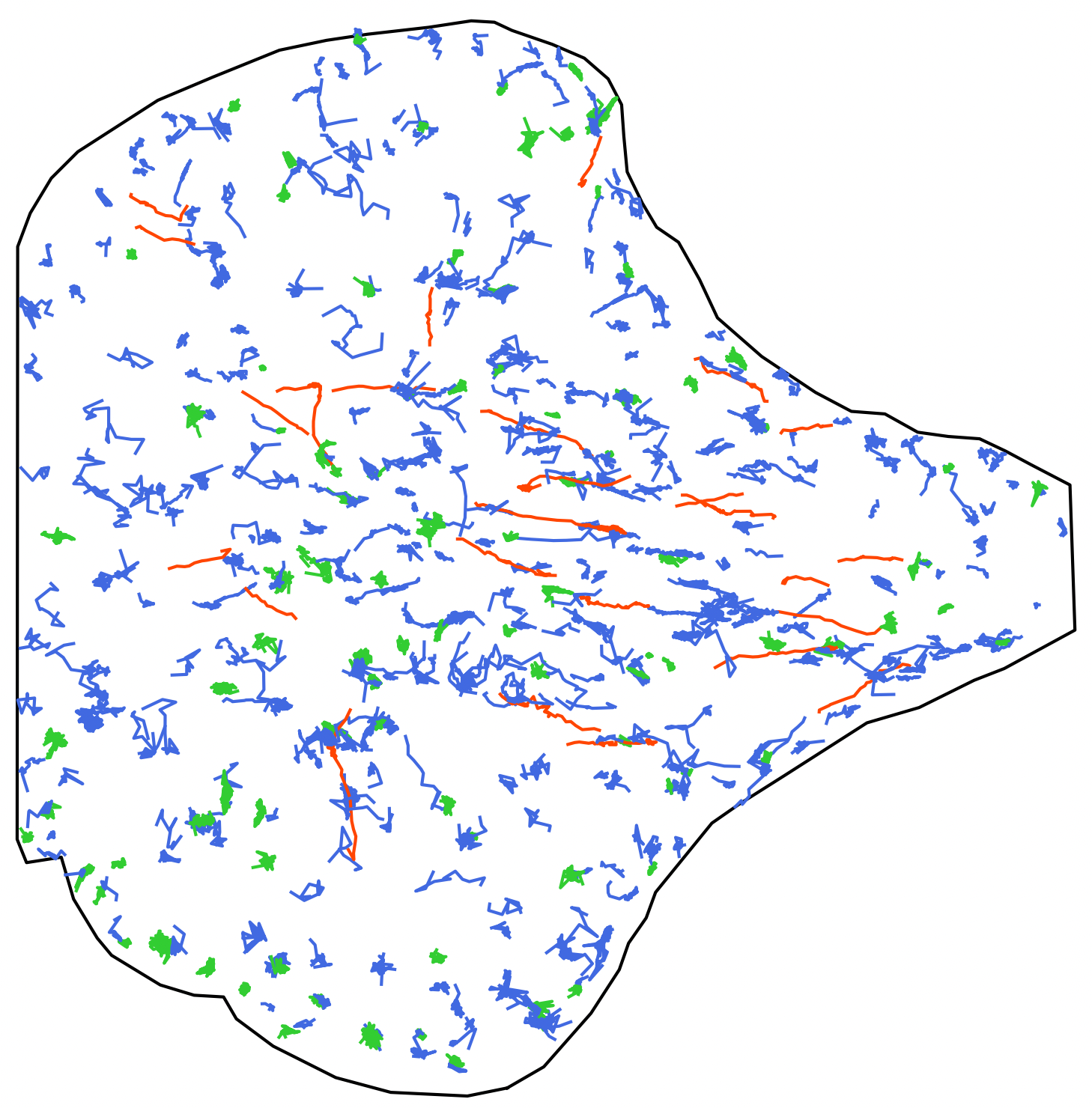} 
     \end{tabular}
    \caption{Top: first frame of the raw sequence showing in bright spots the location of Langerin proteins (on the left) and of Rab11 proteins (on the right). Bottom: Set of all trajectories detected and tracked over the sequence of Langerin proteins (on the left) and Rab11 proteins (on the right), colored by their estimated motion regime (Brownian in blue, superdiffusive in red and subdiffusive in green).}
   \label{real_data}
\end{figure}

The dataset we consider comes from the observation by a 3D multi-angle TIRF (total internal reflection fluorescence) microscopy technique  of the intracellular trafficking of
m-Cherry Langerin and GFP Rab11 proteins in a RPE1 living cell \cite{Boulanger2014}, specifically projected along the z-axis onto the 2D plane close to the plasma membrane.  This provides a 2D image sequence of 1199 frames, each lasting 140 ms and showing the simultaneous locations of the two types of proteins. The two images at the top of Figure~\ref{real_data} depict the first frame of the raw sequence for the Langerin fluorescent channel and the Rab11 fluorescent channel, respectively, recorded simultaneously using a dual-view optical device.  Note that the cell adheres on a fibronectin micropattern, which constrains intracellular constituents such as cytoskeleton elements and gives a reproducible shape, explaining the ``umbrella'' shape of the cell. 
These raw sequences are post-processed following \cite{pecot2008patch,pecot2014background}, then each bright spot is represented by a single point, and we apply the  U-track algorithm \cite{jaqaman2008} to estimate particle trajectories. The bottom images of Figure~\ref{real_data} show the  resulting  trajectories for the  Langerin channel and the Rab11 channel, respectively. These trajectories have been further analysed by the method developed in \cite{Briane18} to classify them into three diffusion regimes: Brownian, superdiffusive and subdiffusive, which corresponds to the blue, red and green colors, respectively,  in Figure~\ref{real_data}.

To be more specific in the analysis of all trajectories, we fit three parametric models to each of them, following \cite{Briane18}, depending on their regime:
\begin{enumerate}[itemsep=0pt]
\item For a Brownian regime (in blue): a Brownian motion, 
\item for a superdiffusive regime (in red): a Brownian motion with constant drift,
\item for a subdiffusive regime (in green): an Ornstein-Uhlenbeck process. 
\end{enumerate}
 Each trajectory has its individual parameters (see Examples~\ref{brownian} and \ref{super sub}), estimated by maximum likelihood. Furthermore, some trajectories may change from one regime to another, which corresponds  to a ``transformation'' in the BDM model that will be specified in the next section. 

\begin{figure}[h]
     \centering
       \includegraphics[height=4.8cm]{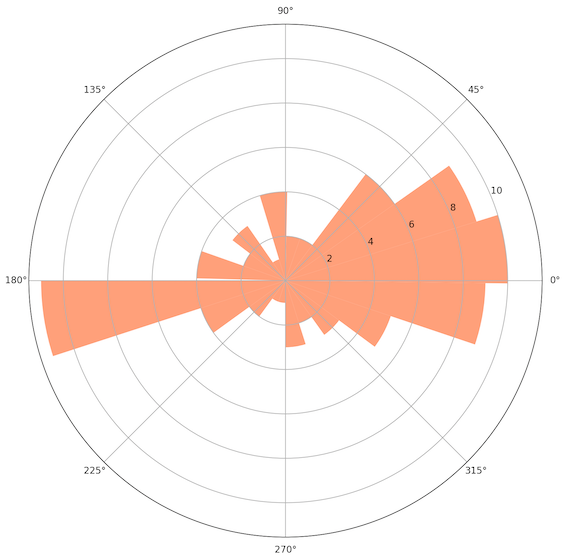}
       \hspace{30pt}
       \includegraphics[height=4.5cm]{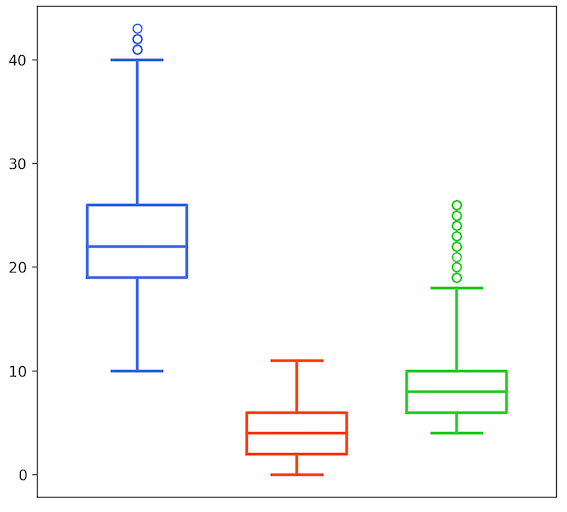}

       \vspace{0.4cm}

           \includegraphics[height=4.5cm]{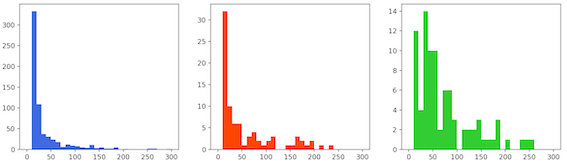}
      \caption{Descriptors of the Langerin trajectories of the real-data sequence. Top-left: circular histogram of the deviation angle (from the direction towards the center of the cell) of the drifts of the superdiffusive trajectories. Top-right: boxplots of the number of trajectories per frame, according to their regime (blue: Brownian, red: superdiffusive, green: subdiffusive). Bottom: histograms of the lifetime (in frames) of each trajectory according to its regime (same color label).}
   \label{figlongtraj}
\end{figure}

Figure~\ref{figlongtraj} summarises different features of the obtained trajectories for the Langerin sequence (the same characteristics have been analysed for the Rab11 sequence, but are not detailed here). 
The histograms at the bottom display the duration of all trajectories (in frames), according to their regime. We can observe that the (blue) Brownian  and (red) superdiffusive trajectories have quite a short lifetime in average, in comparison with the subdiffusive trajectories (in green). The top-right boxplots represent the distribution of the number of particles per frame, according to their regime: there is a majority of Brownian motions, followed by subdiffusive motions and a minority of superdiffusive motions. Finally, 
the top-left circular histogram aims at depicting the orientation of the drift vectors for the superdiffusive (red) trajectories. Specifically, for this plot, we have recorded the deviation of the drift angle (in degrees) with respect to the direction towards the center of the cell. For instance this deviation is $0^\circ$ if the drift goes towards the center, and $180^\circ$ if it goes in the opposite direction. It appears from this plot that most deviation angles are around $0^\circ$ or  $180^\circ$, meaning that the red trajectories mainly move in a radial direction going to (or starting from) the center of the cell.  

The above descriptors will be helpful to calibrate the parameters of the BDM model in the next section and they will also serve as benchmarks to evaluate the quality of our simulations. However it is important to keep in mind that they come with some approximations and errors induced by imperfect tracking algorithms. In particular, no trajectory can last less than 10 frames in the data, which is a minimal length of detection for our tracking method. It is also clear in the bottom plots of Figure~\ref{real_data}, that some directed (superdiffusive) trajectories appear wrongly in blue, which can be explained by the multiple testing procedure of \cite{Briane18} that aims at minimizing the number of false positives (that are bad green or bad red trajectories) to the detriment of possibly too many false negatives (that are wrong blue trajectories). 

\medskip

Concerning the births and deaths of trajectories, we summarize in Table~\ref{tabnbbirthdeath} their total numbers observed in the real-dataset, according to the type of proteins and motion regime. The number of regime transformations is in turn given in Table~\ref{tab-transfo} for the Langerin proteins. For the Rab11 proteins, only one switching from a Brownian motion to a subdiffusive motion was observed during the sequence.

\begin{table}
\centering
\renewcommand{\arraystretch}{1.3}
\begin{tabular}{|c|c|c|c|c||c|c|}
 \cline{3-7}
  \multicolumn{2}{c|}{}&Brownian&Superdiffusive&Subdiffusive&   \multicolumn{2}{c|}{Total} \\
  \hline
    \multirow{2}{*}{Births} & Langerin& 603 &78&66& 747 & \multirow{2}{*}{1248}\\ \cline{2-6}
    & Rab11 & 393&24&84& 501 & \\
     \hline 
  \multirow{2}{*}{Deaths} & Langerin & 602 &77&89& 768 & \multirow{2}{*}{1282}\\ \cline{2-6}
    & Rab11 & 395&26&93 & 514 &\\ \hline
\end{tabular}
\caption{Total number of births and deaths of trajectories observed in the realdataset sequence, according to the type of proteins and the motion regime.  }
     \label{tabnbbirthdeath}
\end{table}

\begin{table}
\centering
\renewcommand{\arraystretch}{1.2}
\begin{tabular}{|c|ccc|}
\hline
From $\backslash$ To & Brownian & Superdiffusive&Subdiffusive\\ \hline
Brownian & 0 & 0 & 9 \\
 Superdiffusive & 0 & 0 & 1\\
 Subdiffusive & 5 & 0 & 0\\ \hline
\end{tabular}
\caption{Number of observed regime transformations for the real-data Langerin trajectories}
\label{tab-transfo}
\end{table}

To address in detail these jumps dynamics, we leverage the study carried out for the same dataset in \cite{Lavancier_LeGuevel}, where it has been concluded that for each type of proteins and motion regimes, the birth intensity is constant, like in Example~\ref{constant intensity}, while the death intensity is proportional to the number of existing particles, like in  Example~\ref{cardinality intensity}. Given the small number of observed motion regime transformations,  its intensity can also be considered as constant. Concerning the transition probability functions, the deaths occur uniformly over all existing particles, like in Example~\ref{unif deaths}. As to the birth transition, there is no reason to choose another density than the uniform distribution over the cell for the Rab11 proteins (Example~\ref{unif births}). But due to colocalization (as observed for this dataset in \cite{Lavancier20}), the birth density for the Langerin positive structures can be approximated by a mixture between a uniform distribution, for $93\%$ of the Langerin births, and a colocalized density around the existing Rab11 vesicles, like in Example~\ref{coloc births}, for $7\%$ of the Langerin births. These proportions, along with the other parameters, have been estimated by maximum likelihood.

\subsection{Simulation of synthetic sequences}\label{sec:simus}

\subsubsection{Model parameters setting}
Based on the data analysis of the previous section, we are now in position to specify all characteristics of the BDM process with transformations presented in Section~\ref{sec:model}, so as to mimic the joint dynamics of Langerin/Rab11 proteins within a cell. 
To make the connection with the theoretical notation, the region of interest $\Lambda$ represents the cell in dimension $d=2$. Each particle in $\Lambda$ will be marked by a label from the set 
$\M=\left\{ (L,1),(L,2),(L,3), (R,1),(R,2),(R,3)\right\}$, where $L$ stands for the Langerin proteins, $R$ for the Rab11 proteins, and the number $1,2$ or $3$ indicates the motion regime of the particle: Brownian, superdiffusive or subdiffusive, respectively. 

Concerning the motion of each trajectory, it follows the regime indicated by its mark and is in agreement with the observed trajectories from the real-dataset detailed in the previous section, see also Examples~\ref{brownian} and \ref{super sub}:
\begin{itemize}
 \item For a Brownian motion, we draw the diffusion coefficient according to the empirical distribution of the diffusion coefficients estimated from the Brownian motions of the real-dataset, for the same type of proteins ($L$ or $R$);
 \item For a superdiffusive motion, we generate a Brownian motion with constant drift, with the same strategy for the choice of the diffusion coefficient, and where the drift vector is chosen as follows: it is by default oriented towards the center of the cell, this orientation being subjected to a deviation drawn from the empirical distribution depicted in the top-left circular histogram of Figure~\ref{figlongtraj}. In addition, its norm is drawn from the empirical distribution of the drift norms observed from the real-dataset. Here again, each set of parameters is distinct for the Langerin and Rab11 proteins;
 \item For a subdiffusive motion, we generate an Ornstein-Uhlenbeck process with diffusion coefficient $0.4$ for all particles (which is the average from the real-dataset), and parameter $\lambda=5.96$ for the Langerin proteins, and $\lambda=7.41$ for the Rab11 proteins. 
\end{itemize}
In these values, the unit is pixels, and one pixel is $160\times 160$ nm$^2$ in our images. 

\medskip

Concerning the intensity functions, we set the birth intensity and the transformation intensity to constant values, as concluded from the real-data analysis. In agreement with Table~\ref{tabnbbirthdeath},  the total birth intensity can be estimated by $\beta(\x)=1248/167.86=7.43$, whatever the configuration $\x$ of particles is, because 1248 is the total number of observed births and $167.86$ is the total time length of the sequence (in seconds). Similarly we set $\tau(\x)=16/167.86=0.095$ since 16 transformations have been observed in the real sequence. For the death intensity, for each mark $m\in\M$, we let it proportional to the number of particles, that is $\delta_m n_m(\x)$, where $n_m(\x)$ is the number of particles with mark $m$ in the configuration $\x$ and $\delta_m$ has been estimated from the real-dataset as follows: 
    $\delta_{m} = 0.17$ if $m=(L,1)$, 
    $\delta_{m} =0.14$ if $m=(L,2)$,
    $\delta_{m} =0.07$ if $m=(L,3)$, 
    $\delta_{m} =0.21$ if $m=(R,1)$,
    $\delta_{m} =0.25$ if $m=(R,2)$,
and     $\delta_{m} =0.08$ if $m=(R,3)$. The total death intensity for the configuration $\x$ of particles is then $\delta(\x)=\sum_{m\in\M} \delta_m n_m(\x)$.

Finally we set the transition probability functions as follows. For the death transition, the probability to kill the particle $x_i=(z_i,m_i)$ in the configuration $\x$  is set to 
$$p^\delta((z_i,m_i)|\x)=\frac{\delta_{m_i}}{\delta(\x)},$$
which means that we first draw the mark $m$ with probability $\delta_m n_m(\x) /\delta(\x)$ and then the particle uniformly among all existing particles with mark $m$. For the transformation transition, we first select the type of proteins to transform with probability $15/16$ for Langerin and $1/16$ for Rab11, in line with the transformations rates observed in the real-sequence, second we choose a particle uniformly among all existing particles of this type, and third, as in Example~\ref{indep transfo}, we apply a regime transformation with respect to the following transition matrix (from the regime in rows to the regime in columns):
$$\bordermatrix{ & 1&2&3\cr 
1 & 0& 1/4&3/4 \cr 
2 & 1/2&0 & 1/2 \cr
3 & 1 & 0&0 \cr}.$$
This matrix is in agreement with Table~\ref{tab-transfo} concerning the Langerin proteins, where we have added some possible transitions from regime 1 to 2, and from regime 2 to 1, that appear to us likely to occur, even if they were not observed in the (quite rare) transformations in the real-sequence. The same transition matrix has been set for the Rab11 proteins, since there is not enough observed transformations in the real-sequence (only one) to design a finer choice. 

It remains to set the birth transition probability. First, we select the type of protein to create with probability $747/1248$ for the Langerin proteins and $501/1248$ for the Rab11 proteins, following  Table~\ref{tabnbbirthdeath}. If the selected type is Rab11, then it is generated uniformly in the cell with regime $1$ with probability $0.784$, $2$ with probability $0.048$ and $3$ with probability $0.168$, which corresponds to the relative proportion of births of each regime over all births for the real Rab11 sequence. If the selected type is Langerin, then we flip a coin for colocalization with probability $p=0.07$. If there is no colocalization, then the new Langerin protein is generated uniformly in the cell with 
%regime $1$ with probability \com{$0.766$}, $2$ with probability \com{$0.146$} 
regime $1$ with probability $0.807$, $2$ with probability $0.104$
and $3$ with probability $0.088$ (the observed relative proportions of births). If there is colocalization, then the new Langerin protein is generated around an existing Rab11 protein according to the density \eqref{gaussian mix} in Example~\ref{coloc births}, where by maximum likelihood estimation $\sigma=1.1$. In this case, the regime of the new Langerin protein and its drift vector for a superdiffusive motion are similar as the those of its colocalized Rab11 protein.

\subsubsection{Analysis of resulting simulations}

\begin{figure}[h]
     \centering
       \includegraphics[height=4.8cm]{traj60}
       \hspace{30pt}
       \includegraphics[height=4.5cm]{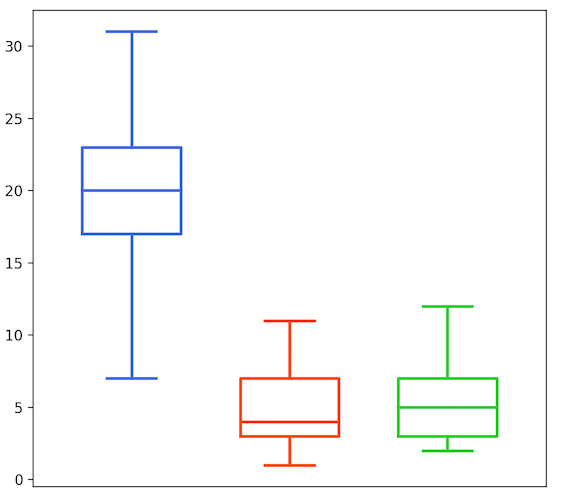}

       \vspace{0.4cm}

           \includegraphics[height=4.5cm]{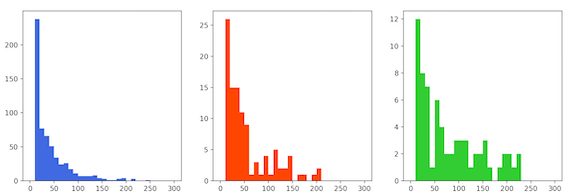}
      \caption{Descriptors of the Langerin trajectories of a first simulated sequence. Top-left: set of trajectories, coloured according to their motion regime (blue: Brownian, red: superdiffusive, green: subdiffusive). Top-right: boxplots of the number of trajectories per frame, according to their regime (same color label). Bottom: histograms of the lifetime (in frames) of each trajectory according to its regime (same color label).}
   \label{simu1}
\end{figure}

\begin{figure}[h]
     \centering
       \includegraphics[height=4.8cm]{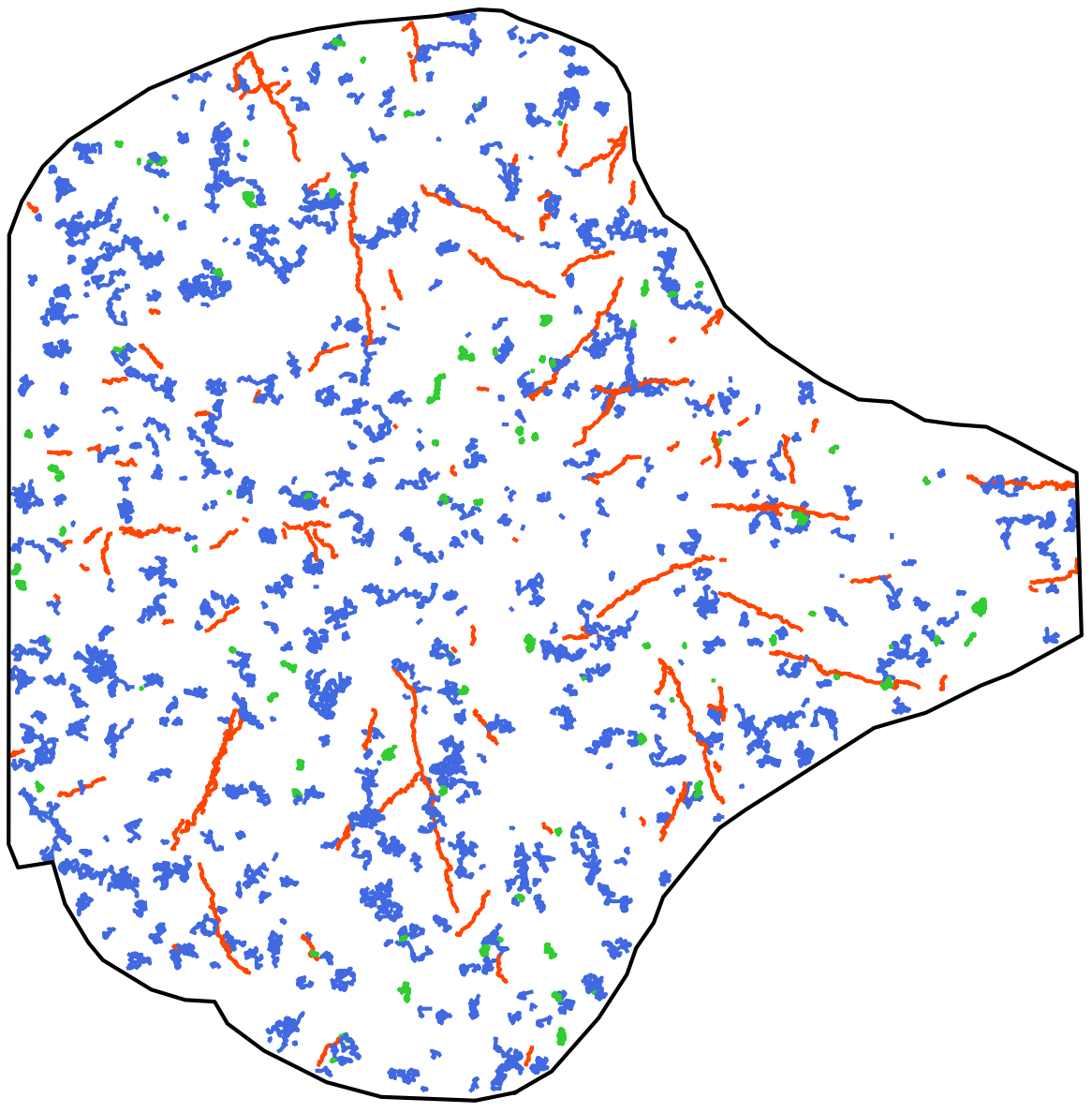}
       \hspace{30pt}
       \includegraphics[height=4.5cm]{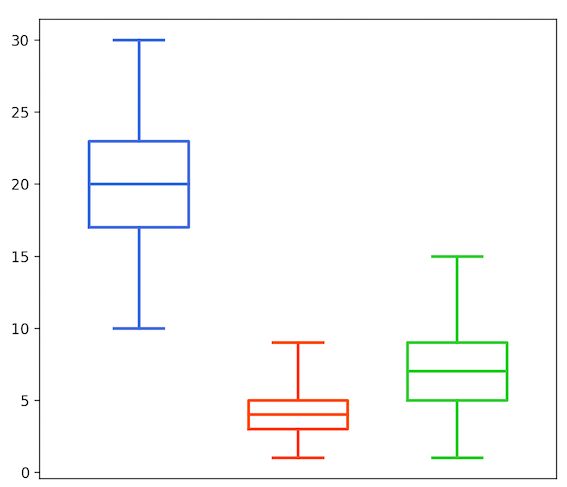}

       \vspace{0.4cm}

           \includegraphics[height=4.5cm]{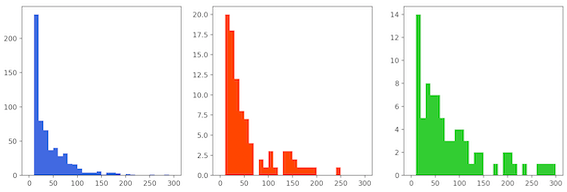}
      \caption{Descriptors of the Langerin trajectories of a second simulated sequence, as in Figure~\ref{simu1}.}
   \label{simu2}
\end{figure}

We have generated 100 sequences following the model of the previous section, during the same time length as the real-sequence of Section~\ref{sec:data}, that is $167.86$s for 1199 frames.  Some descriptors concerning the generated Langerin trajectories coming from two simulated sequences are depicted in Figures~\ref{simu1} and \ref{simu2}, that are to be compared with the similar outputs of  the real-data in Figures~\ref{real_data} and \ref{figlongtraj}. The results for other simulated sequences can be seen in our GitHub repository. We have also summarized the mean number of births and deaths over the 100 simulated sequences in Table~\ref{tab-simu}, to be compared with Table~\ref{tabnbbirthdeath}.  
Both graphical and quantitative results demonstrate that our generator is able to create a joint dynamics with comparable features as those observed in the real-data sequence.

\begin{table}
\centering
\renewcommand{\arraystretch}{1.3}
\begin{tabular}{|c|c|c|c|c||c|c|}
 \cline{3-7}
  \multicolumn{2}{c|}{}&Brownian&Superdiffusive&Subdiffusive& \multicolumn{2}{c|}{Total} \\
  \hline
    \multirow{2}{*}{Birth} & Langerin& 581 &85 &76 & 742 &  \multirow{2}{*}{1239}\\ \cline{2-6}
    & Rab11 & 391&22&83& 496 &\\
     \hline
  \multirow{2}{*}{Death} & Langerin & 584&86&77& 747 &\multirow{2}{*}{1248}\\ \cline{2-6}
    & Rab11 & 397&23&80 & 500 & \\ \hline
\end{tabular}
\caption{Mean total number of births and deaths of trajectories per sequence, over 100 simulated sequences.}     \label{tab-simu}
\end{table}

\section{Conclusion and perspectives}

We have leveraged an original stochastic model, namely a multitype birth-death-move process with transformations, in order to generate realistic image sequences of biomolecules dynamics within a cell. 
This stochastic process not only models the individual trajectory of particles, but it is also able to generate the appearance (i.e., birth), disappearance (i.e., death) and regime switching (i.e., transformation) of each trajectory over time, the occurence of which possibly depending on the configuration of the existing particles (e.g. their numbers). The model is very flexible and is specified thanks to three sets of parameters: 1) a system describing the set of trajectories (typically a system of stochastic differential equations); 2) the intensity functions, ruling the waiting time before a new appearance, a disappearance or a switching; 3) the transition probability functions, driving where a new particle appears when there is a birth, which particle disappears when there is a death, and which particle switches its regime (and how) when there is a transformation. Numerous examples of these model specifications have been detailed. We demonstrated the relevance of this approach by generating realistic image sequences of the joint dynamics of  Langerin/Rab11 proteins within a cell, based on a preliminary data-based analysis in order to finely calibrate the model.

Since the model is very flexible, an important step is the choice of model characteristics and parameters. In order to improve our approach, a deeper empirical study based from many image sequences might help calibrating robustly the model.  Once the parameters are fitted, the generation of an image sequence is quite fast: about one minute on an regular laptop for the generation of 2000 frames containing each about 70 trajectories in interaction. In an effort to generate even more realistic image sequences, we may consider to blur the system of generated particles using the point spread function, and to add some noise and background. In relation, additional features could be computed from both the real-image sequence and the synthetic ones in order to strengthen the quality assessment  of the generator.

\appendix
\section*{Appendix: Algorithm for simulation}

We provide in this appendix a formal algorithm to simulate a birth-death-move process with mutations (or transformations), following the construction of Section~\ref{sec:algo}. It is a refinement of the algorithm introduced in \cite{Lavancier_LeGuevel} for a birth-death-move process (without transformations). 
The idea is to generate the inter-jump move on a small time length $\Delta$, then to test whether a jump has occurred during this period (this is with probability $p$ in the following Algorithm~\ref{algo_simu_fast}). If so, we generate the jump time and the jump. If not, we continue the simulation of the inter-jump move on a further time length $\Delta$,  test whether a jump has occurred, and so on. The algorithm is valid  whatever $\Delta>0$ is, but an efficient choice is to  set a small value for $\Delta$. Its implementation is available in our GitHub repository. 

 We let as in Section~\ref{sec:algo} $\alpha=\beta+\delta+\tau$ and $T_0=0$. We denote in Algorithm~\ref{algo_birth-death} $\X_{T_j^-}$ the configuration just before the jump time $T_j$.
In order to run Algorithms~\ref{algo_simu_fast} and \ref{algo_birth-death}, we need the following inputs:
\begin{itemize}
\item $T>0$: final time of simulation;

\item $\X_0\in E$: initial configuration of particles;

\item $\Delta>0$: small time length for piecewise simulation;

\item $\beta$, $\delta$, $\tau$: intensity functions of births, deaths and transformations;

\item $p^\beta$, $p^\delta$, $p^\tau$: transition probability for a birth, a death and a transformation;

\item $Algo.Move(\mathbf y_0,\Delta)$: algorithm that returns, for $\mathbf y_0\in E$ and $\Delta>0$,  $n(\mathbf y_0)$  trajectories on $[0,\Delta]$ following the system of SDEs $(Move)$ with initial configuration $\mathbf y_0$.

\end{itemize}

\begin{algorithm}
\caption{Simulation on the time interval $[0,T]$}
\label{algo_simu_fast}
 {\bf set}  $t=0$ and $j=0$ \;
 \While{$t<T$}{
% {\bf set} $\Delta_j=\min(\Delta,T-T_j)$, $k=1$ and  $p_{0}=1$ \;
  {\bf set} $k=1$  \;
%$k=1$ \;
%$p_{0}=1$ \;
 \While {$k \neq 0$ }{
  {\bf set} $\Delta_j=\min(\Delta,T-T_j-(k-1)\Delta)$ \;

% {\bf generate}  $(\Y_s)_{0\leq s\leq \Delta_j-(k-1) \Delta}$ by Algorithm $Move(\X_{T_j}, \Delta_j-(k-1) \Delta)$  \;
 {\bf generate}  $(\Y_s)_{0\leq s\leq \Delta_j}$ by  $Algo.Move(\X_{T_j+(k-1) \Delta}, \Delta_j)$  \;
% {\bf set}  $p_k=\exp\left( -  \int_0^{\Delta_j} \alpha( \Y_u)du \right)/p_{k-1}$ \;
 {\bf set}  $p=\exp\left( -  \int_0^{\Delta_j} \alpha( \Y_u )du \right)$ \;
 {\bf generate} $U\sim U([0,1])$\;
  \eIf{$U\leq p$} {
 {\bf set}  $\X_s=\Y_{s-T_{j}-(k-1) \Delta}$ for $s\in [T_{j}+(k-1) \Delta,T_{j}+(k-1) \Delta + \Delta_j]$ \;
 
   \eIf {$\Delta_j=T-T_j-(k-1)\Delta$}{
   	$k=0$ \;
	$t \leftarrow T$ \;}{
   $k \leftarrow k+1$ \;
}
}{

%   {\bf generate} the waiting time $\tau_j$ before the next jump according to the distribution
%       \begin{multline*}\forall s\in [(k-1)\Delta,  \Delta_j],\quad \P\left(\tau_{j}< s | (k-1)\Delta<\tau_j <  \Delta_j\right)\\
%       =\frac 1 {p_{k-1}(1-p_k)} \left(\exp\left( - \int_0^{(k-1)\Delta} \alpha( Y_u)du \right) -\exp\left( - \int_0^{s} \alpha( Y_u)du \right)\right)\end{multline*} 
          {\bf generate} the waiting time $\tau$ according to the distribution
%    \begin{multline*}\forall s\in [(k-1)\Delta,(k-1)\Delta+  \Delta_j],\\
%    \P\left(\tau< s | (k-1)\Delta<\tau < (k-1)\Delta+ \Delta_j\right)
%       =\frac  { 1 -\exp\left( - \int_0^{s-(k-1)\Delta} \alpha( Y_u)du \right)}{(1-p_k)} \end{multline*} 
  \begin{multline*}
    \P\left(\tau< s | (k-1)\Delta<\tau < (k-1)\Delta+ \Delta_j\right)\\
       =\frac  { 1 -\exp\left( - \int_0^{s-(k-1)\Delta} \alpha( Y_u)du \right)}{(1-p)}, \quad \forall s\in [(k-1)\Delta,(k-1)\Delta+  \Delta_j]\, ; \end{multline*} 

{\bf set} $T_{j+1}=T_j+\tau$ and $\X_s=\Y_{s-T_{j}-(k-1) \Delta}$ for $s\in [T_{j}+(k-1) \Delta,T_{j+1}[$ \;
    {\bf generate} $\X_{T_{j+1}}$ by Algorithm~\ref{algo_birth-death} given $\X_{T_{j+1}^-}=\Y_{T_{j+1}-T_{j}-(k-1) \Delta}$ \;
      $k=0$ \;
   $t\leftarrow T_{j+1}$ \; 
   $j\leftarrow j+1$ \;
     
}
}
   
 }
 \end{algorithm}

\begin{algorithm}
\caption{Simulation of the jump at $t=T_{j+1}$ given $\X_{T_{j+1}^-}$}\label{algo_birth-death}
 {\bf generate} $U\sim U([0,1])$ \;
 \eIf{$U\leq \beta(\X_{T_{j+1}^-})/\alpha(\X_{T_{j+1}^-})$}  {
{\bf generate}  a new particle $(z,m)$ according to $p^\beta((z,m)|\X_{T_{j+1}^-})$\;
{\bf set} $\X_{T_{j+1}}=\X_{T_{j+1}^-}\cup (z,m)$\;
}{ 
\eIf{$U\leq (\beta+\delta)(\X_{T_{j+1}^-})/\alpha(\X_{T_{j+1}^-})$}{
{\bf sample} $x_i\in\X_{T_{j+1}^-}$ according to $p^\delta(x_i|\X_{T_{j+1}^-})$\;
{\bf set} $\X_{T_{j+1}}=\X_{T_{j+1}^-}\setminus (z,m)$\;}{
{\bf sample}  $x_i=(z_i,m_i)\in\X_{T_{j+1}^-}$ and $m\in\M$  according to  $p^\tau((z_i,m_i),m|\X_{T_{j+1}^-})$\;
{\bf set} $\X_{T_{j+1}}=(\X_{T_{j+1}^-}\setminus (z_i,m_i))\cup (z_i,m)$\;
}}
  \end{algorithm}

 \bibliographystyle{acm} 
\bibliography{bibref}

\end{document}